\documentclass[12pt,preprint]{aastex}
\usepackage{natbib,graphicx}
\usepackage{epstopdf}                                      
\usepackage[breaklinks,colorlinks,urlcolor=blue,citecolor=blue,linkcolor=blue]{hyperref} 
\graphicspath{{./}}

\newcommand{\ldl}{$\lambda/{\Delta}{\lambda}$}

\newcommand{\kms}{km~s$^{-1}$}
\newcommand{\masyr}{mas~yr$^{-1}$}

\begin{document}

\title{Properties of the Nearby Brown Dwarf WISEP J180026.60+013453.1}

\author{John E. Gizis\altaffilmark{1}, 
Adam J. Burgasser\altaffilmark{2},
Frederick J. Vrba\altaffilmark{3}
}

\altaffiltext{1}{Department of Physics and Astronomy, University of Delaware, Newark, DE 19716, USA}
\altaffiltext{2}{Center for Astrophysics and Space Science, University of California San Diego, La Jolla, CA 92093, USA}
\altaffiltext{3}{US Naval Observatory, Flagstaff Station, 10391 West Naval Observatory Road, Flagstaff, AZ 86001, USA}

\begin{abstract}
We present new spectroscopy and astrometry to characterize the nearby brown dwarf WISEP J180026.60+013453.1. The optical spectral type, L7.5, is in agreement with the previously reported near-infrared spectral type. The preliminary trigonometric parallax places it at a distance of $8.01 \pm 0.21$ pc, confirming that it is the fourth closest known late-L (L7-L9) dwarf.  The measured luminosity, our detection of lithium, and the lack of low surface gravity indicators indicates that WISEP J180026.60+013453.1 has a mass $0.03 < M < 0.06 M_\odot$ and an age between 300 million and 1.5 billion years according to theoretical substellar evolution models. The low space motion is consistent with this young age. We have measured the rotational broadening ($v \sin i = 13.5 \pm 0.5$ \kms), and use it to estimate a maximum rotation period of 9.3 hr.  
\end{abstract}


\section{Introduction\label{intro}}

The closest stars to the Sun make up a uniquely valuable sample for astronomical studies. Nearby L dwarfs are especially significant due to their relative rarity and low intrinsic luminosities.  \citet{2012ApJ...753..156K} list 243 stars and brown dwarfs in 182 systems within 8 pc of the Sun, including just three L dwarfs. Discoveries of additional nearby very-low-mass stars and brown dwarfs \citep{2013ApJ...767L...1L,2013A&A...557A..43B,2014A&A...567A...6P,2014ApJ...786L..18L,2014AJ....147..113C,2014A&A...561A.113S} have continued. Of these, WISE J104915.57-531906.1A (Luhman-16A) is a late-L dwarf at $\sim 2$ pc \citep{2013ApJ...767L...1L,2013ApJ...770..124K,2013ApJ...772..129B}.  Additionally, WISE J072003.20-084651.2  \citep{2014A&A...561A.113S} is a low-mass binary system with a combined-light optical classification of 
L$0 \pm1$ \citep{2015A&A...574A..64I} and component near-infrared classifications of M9.5+T5 \citep{2015AJ....149..104B}. Although currently 6 pc distant, it likely passed within 0.25 pc from the Sun in the last 70 kyr \citep{2015ApJ...800L..17M}.  

WISEP J180026.60+013453.1 (hereafter W1800+0134) was discovered by \citet{2011AJ....142..171G} using the Wide-field Infrared Survey Explorer \citep{2010AJ....140.1868W} and Two Micron All-Sky Survey \citep{2mass}. We classified W1800+0134 as an L7.5 dwarf in the near-infrared and estimated its distance as $8.8 \pm 1.0$ pc.  In this paper, we present additional observations and analysis of this brown dwarf, including a preliminary trigonometric parallax that places it at $8.01 \pm 0.21$ pc, confirming that it is the fourth nearest late-L (L7-L9) dwarf.

\section{Observations}

\subsection{Spectroscopy}

\begin{figure}
\plotone{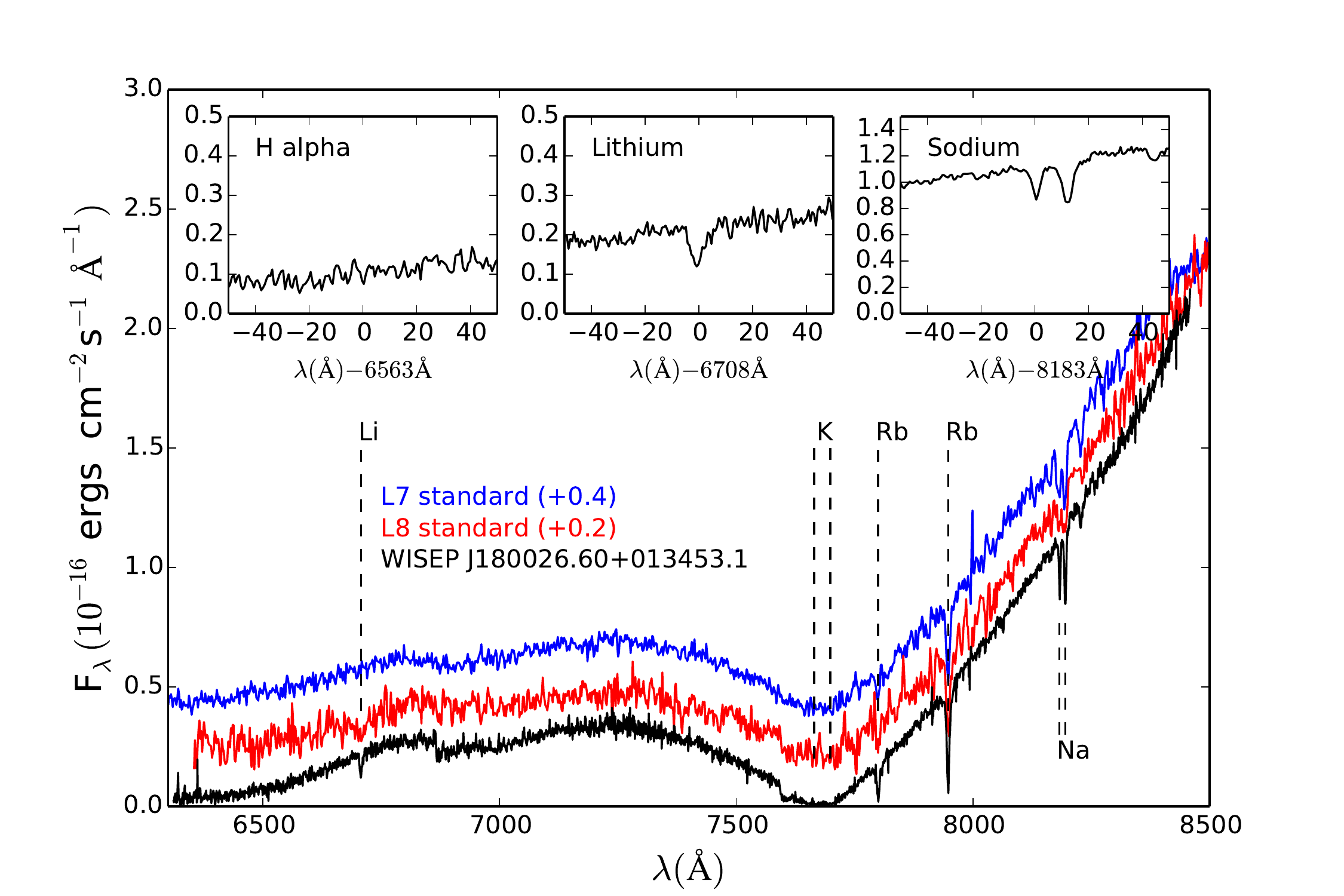}
\caption{The optical spectrum is consistent with a type of L7.5. Note the detection of lithium in absorption and the absence of any H$\alpha$ feature. Also shown are the
\citet{1999ApJ...519..802K} spectra of the L7 dwarf standard DENIS-P J0205.4-1159 (blue, scaled by a factor 1.51 to match W1800+0134) and the L8 dwarf standard 2MASSW J1632291+190441 (red, scaled by a factor of 7.6). Each standard is offset vertically to allow comparison, but we caution that the \citet{1999ApJ...519..802K} spectra were taken with lower resolution which affects the appearance of the narrow atomic lines. \label{fig-gemini}}
\end{figure}

Optical (far-red) long-slit spectra were obtained with the Gemini-North telescope (Gemini program GN-2012B-Q-105) on UT Date 14 September 2012 with the GMOS spectrograph \citep{Hook:2004lr} using grating R831. Four 600 second exposures were taken during cloudy conditions. The wavelength coverage was 6340 to 8460\AA~with a resolution of $\sim2$\AA. The seeing (full width at half maximum) was 0.64 arcsec, and W1800+0134 appears single in the $z$-band finder image.  The spectrum was reduced with the Gemini GMOS IRAF package and is shown in Figure~\ref{fig-gemini}.  The signal-to-noise ranges from $\sim8$ at the blue end to  $\sim80$ at the red end.  Comparing to the \citet{1999ApJ...519..802K} spectral standards, we find that the spectral type is L$7.5 \pm 0.5$, midway between the L7 and L8 standards. We caution that our spectrum does not include the 8460\AA~ to 10000\AA~ region necessary for a definitive optical spectral type, but it is consistent with the near-infrared L7.5 spectral type reported in the discovery paper.  Lithium absorption is present with an equivalent width (EW) of 2.7 \AA. No H$\alpha$ emission or absorption is detected (EW $< 0.5$ \AA).  

\begin{figure}
\plotone{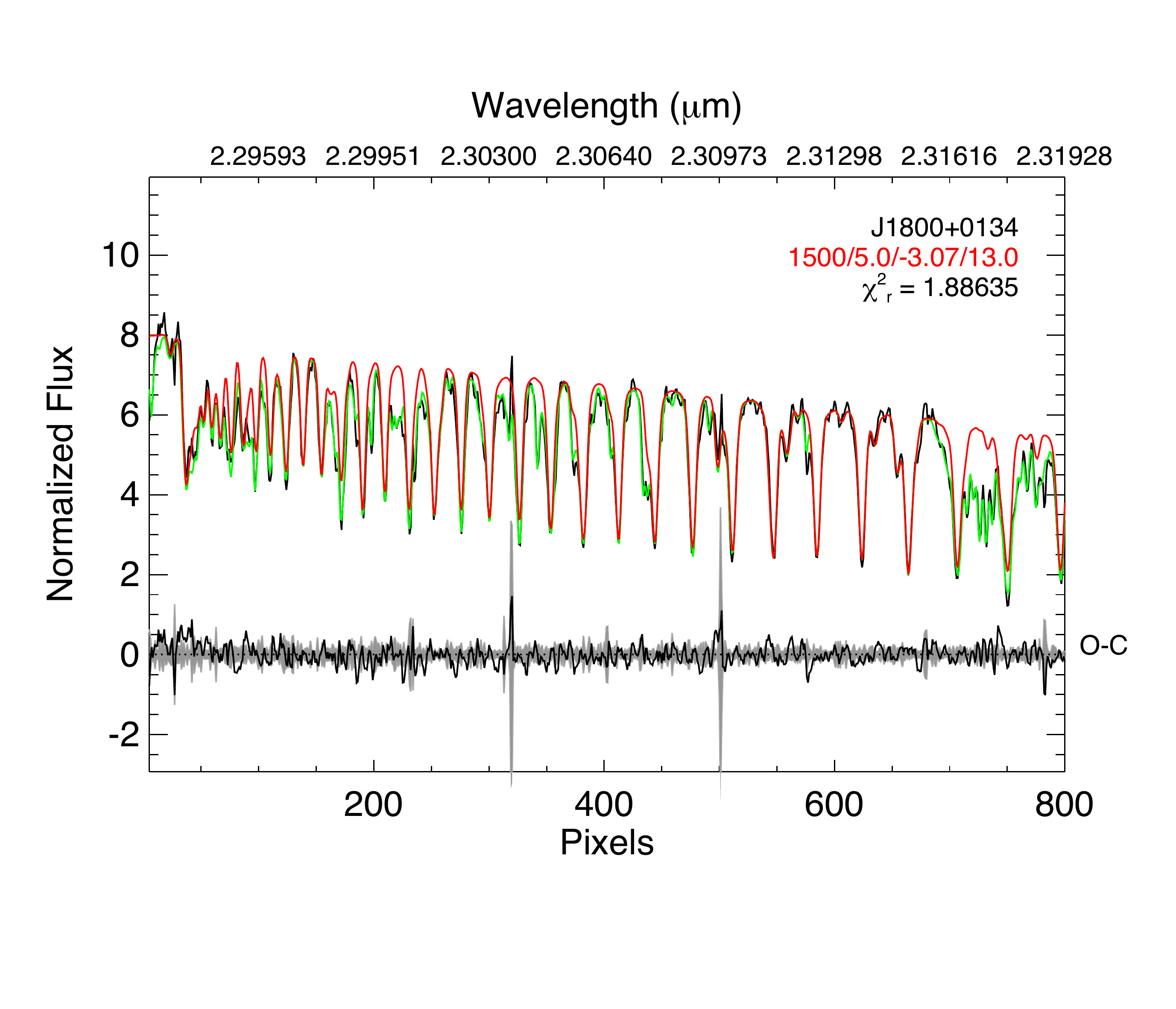}
\caption{The NIRSPEC spectrum of WISEP J180026.60+013453.1 with the best fit model. The data are shown in black, the model of the intrinsic spectrum in red, and the model including telluric absorption is in green. The $v_{rad}$ and $v \sin i$ reported in Table~\ref{tabproperties} are the best-fit values from our MCMC analysis.  
\label{fig-keck}}
\end{figure}

W1800+0134 was observed in clear and dry conditions on UT Date 6 July 2011 with the Keck II NIRSPEC near-infrared echelle spectrograph  \citep{McLean:2000lr}, as part of an ongoing search for radial velocity variables among nearby  L dwarfs \citep{Burgasser:2012qy}.   The source was observed using the high-dispersion mode, N7 filter and 0$\farcs$432$\times$12$\arcsec$ slit to obtain 2.00--2.39~$\mu$m spectra over orders 32--38 with {\ldl} = 20,000 ($\Delta{v}$ = 15~{\kms}) and dispersion of 0.315~{\AA}~pixel$^{-1}$.  The two 120~s  exposures were obtained in nods separated by 7$\arcsec$ along the slit, and a nearby A0 star was observed for telluric calibration.  The resulting near-infrared spectrum was reduced using an adaptation of the REDSPEC package (described in \citealt{2007ApJ...658.1217M}), then compared to a family of theoretic BT-Settl models 
\citep{2014IAUS..299..271A} using a Markov-Chain Monte Carlo (MCMC)   adaption of the forward-modeling method described in \citet{Blake:2010gf} and 
\citet{2015AJ....149..104B}.  The Solar atlas of \citet{Livingston:1991fj} was used to model telluric absorption, and both radial shift and rotational line broadening were simulated in the theoretical models. Figure~\ref{fig-keck} shows the best-fit model from our MCMC chain. Marginalizing over all fits, we infer a radial velocity of $v_{\mathrm{rad}} = -5.1 \pm 0.7$ \kms and projected equatorial rotational velocity $v \sin i = 13.5 \pm 0.5$ \kms.

\subsection{Astrometry}

W1800+0134 was added to the U.S. Naval Observatory's near-infrared parallax program 
 \citep{2004AJ....127.2948V} in July 2011. Astrometric observations were obtained in
J-band using the reconstructed ASTROCAM camera, the optically identical
original of which is described by \citet{Fischer:2003fj}. The preliminary results
presented here are derived from data obtained through May of 2015 and consist of 96
frames obtained on 31 nights. The relative trigonometric parallax is $122.45 \pm 3.18$ mas
with a proper motion of $424.7 \pm  2.1$ \masyr at position angle $153.8 \pm 0.2$
degrees. The correction to absolute parallax for the 14 reference frame stars
was determined to be $2.44 \pm 0.62$ mas based on 2MASS photometry and the
homogenized infrared colors of \citet{1995ApJS..101..117K}; giving an absolute
parallax of $124.89 \pm 3.24$ mas. This object continues on the USNO parallax
program and we expect to obtain a considerably improved parallax based on a
full analysis of the final data set in approximately three years.  
We obtain the luminosity by applying a K-band bolometric correction of $3.19 \pm 0.07$ 
mags \citep{2004AJ....127.3516G,Cushing:2006fk} and the $U$, $V$ and $W$ components of 
space motion from the astrometry and radial velocity. The measurements are listed in 
Table~\ref{tabproperties}. 

\section{Discussion}

All late-L dwarfs are brown dwarfs below the hydrogen-burning limit, but our observations further constrain the properties of W1800+0134. The presence of lithium requires this source to have a mass less than $0.06 M_\odot$ \citep{1993ApJ...404L..17M}. Theoretical evolutionary models predict that an object with luminosity $\log L/L_\odot = -4.5$ with lithium must be $\sim 1.5$ billion years old or younger \citep{1997ApJ...491..856B,2003A&A...402..701B,2008ApJ...689.1327S}. W1800+0136 does not show any signs of low surface gravity in its near-infrared spectrum, and it lies within the field dwarf locus in the \citet{Dupuy:2012fk} near-infrared color-magnitude  diagrams. These imply that $\log g \gtrsim 4.7$ and therefore the age is $\gtrsim 300$ Myr and mass $\gtrsim 0.03 M_\odot$ according to the same evolutionary models. The space motion is consistent with the young thin disk, and W1800+0134 has a 100\% chance of being a field object and a 0\% chance of belonging to any known moving group according to the model of \citet{2014ApJ...783..121G}. As discussed in the discovery paper, there is no spectroscopic evidence of an unresolved companion; in \citet{2015A&A...578A...1H}'s survey, W1800+0134 did not show any companion down to 0.3 arcseconds in the H-band.  The apparent magnitudes and parallax are consistent with a single L7.5 dwarf.  The rotational period should be $\sim 9.3$ hours or less, based on the model predictions of a radius of $\sim 0.10 R_\odot$ and our measurement of $v \sin i = 13.5 \pm 0.5$ \kms.   

Overall, W1800+0134 is a typical L7.5 dwarf. With its relatively large parallax and position near the celestial equator, it is well suited for additional follow-up and detailed study by ground-based telescopes in either hemisphere. For example, its classification at the start of the L dwarf/T dwarf transition, proximity, and relatively dense surrounding starfield, makes W1800+0134 a potentially interesting target for variability studies. While its potentially long rotation period may challenge experimental designs, \citet{2013ApJ...768..121A} and \citet{2014ApJ...785...48B} have argued that such long-period rotators may have larger spot features, and hence more pronounced variability, if spot size correlates with the atmospheric Rhines scale, which is itself proportional to the rotation period. Indeed, the highly variable T1.5 dwarf 2MASS J21392676+0220226, which exhibits $\sim$30\% variability amplitude in the infrared, has a rotation period of 7.7 hr \citep{Radigan:2012vn,Khandrika:2013fk}.  
Based on the observed parallax, W1800+0134 is a confirmed member of the 10 pc sample and may be a member of the 8 pc sample. Assuming a normal distribution for the uncertainty, the probability that W1800+0134 has a distance less than 8.0 pc is 48.6\% if no priors are applied, or 46.6\% if we apply our prior belief that the probability should be proportional to the volume sampled.  Until a much more precise parallax is available, W1800+0134 could be included with this weight in statistical studies of the 8 pc sample such as \citet{2012ApJ...753..156K}.

\begin{deluxetable}{lll}
\tablewidth{0pc}
\tabletypesize{\footnotesize}
\tablenum{1}
\tablecaption{WISEP J180026.60+013453.1}
\tablehead{
\colhead{Parameter} & 
\colhead{W1800+0134} }
\startdata
Sp Type (Optical) & L$7.5 \pm 0.5$\\
$v_{rad}$ \kms & $-5.1 \pm 0.7$ \\
$\pi_{rel} (mas)$ & $122.45 \pm 3.18$ \\
$\pi_{abs} (mas)$ & $124.89 \pm 3.24$ \\
$\mu$ (mas/yr) & $424.7 \pm 2.1$ \\
$\theta$ (deg) & $ 153.79 \pm 0.14$ \\
distance (pc) & $8.01 \pm 0.21$ \\
$U$ \kms & $2.6 \pm 0.6$ \\
$V$ \kms & $-9.5 \pm 0.4$ \\
$W$ \kms & $-13.8 \pm 0.4$ \\ 
$M_J$ (2MASS) & $14.78\pm 0.07$ \\
$M_H$ (2MASS) & $13.60 \pm 0.07$ \\
$M_{Ks}$ (2MASS) & $12.90 \pm 0.06$ \\
$\log L/L_\odot$   & $-4.53 \pm 0.04$ \\
Age\tablenotemark{a} & 300-1500 Myr \\
Mass\tablenotemark{a} & 0.03 - $0.06 M_\odot$  \\ 
\enddata
\tablenotetext{a}{Model dependent, see text}
\label{tabproperties}
\end{deluxetable}

\acknowledgments

We wish to recognize C. Luginbuhl, J. Munn, and T. Tilleman for their many contributions to the USNO infrared astrometry program. 
We thank Radostin Kurtev for helpful discussions of the \citet{2015A&A...578A...1H} imaging survey.  
This work is based in part on observations obtained at the Gemini Observatory, which is operated by the Association of Universities for Research in Astronomy (AURA) under a cooperative agreement with the NSF on behalf of the Gemini partnership: the National Science Foundation (United States), the Science and Technology Facilities Council (United Kingdom), the National Research Council (Canada), CONICYT (Chile), the Australian Research Council (Australia), CNPq (Brazil) and CONICET (Argentina). Some of the data presented herein were obtained at the W. M. Keck Observatory, which is operated as a scientific partnership among the California Institute of Technology, the University of California and the National Aeronautics and Space Administration. The Observatory was made possible by the generous financial support of the W. M. Keck Foundation. The authors wish to recognize and acknowledge the very significant cultural role and reverence that the summit of Mauna Kea has always had within the indigenous Hawaiian community. We are most fortunate to have the opportunity to conduct observations from this mountain.  This research has made use of NASA's Astrophysics Data System, IRAF, and Astropy, a community-developed core Python package for Astronomy \citep{2013A&A...558A..33A}. IRAF is distributed by the National Optical Astronomy Observatory, which is operated by the Association of Universities for Research in Astronomy (AURA) under cooperative agreement with the National Science Foundation.

\bibliographystyle{yahapj}
\bibliography{../astrobib}

\begin{thebibliography}{}
\providecommand\natexlab[1]{#1}
\providecommand\JournalTitle[1]{#1}

\bibitem[{{Allard}(2014)}]{2014IAUS..299..271A}
{Allard}, F. 2014, \href{http://dx.doi.org/10.1017/S1743921313008545}{in IAU
  Symposium, Vol. 299, IAU Symposium, ed. M.~{Booth}, B.~C. {Matthews}, \&
  J.~R. {Graham}}, 271

\bibitem[{{Apai} {et~al.}(2013){Apai}, {Radigan}, {Buenzli}, {Burrows}, {Reid},
  \& {Jayawardhana}}]{2013ApJ...768..121A}
{Apai}, D., {Radigan}, J., {Buenzli}, E., {et~al.} 2013,
  \href{http://dx.doi.org/10.1088/0004-637X/768/2/121}{\JournalTitle{\apj},
  768, 121}

\bibitem[{{Astropy Collaboration} {et~al.}(2013){Astropy Collaboration},
  {Robitaille}, {Tollerud}, {Greenfield}, {Droettboom}, {Bray}, {Aldcroft},
  {Davis}, {Ginsburg}, {Price-Whelan}, {Kerzendorf}, {Conley}, {Crighton},
  {Barbary}, {Muna}, {Ferguson}, {Grollier}, {Parikh}, {Nair}, {Unther},
  {Deil}, {Woillez}, {Conseil}, {Kramer}, {Turner}, {Singer}, {Fox}, {Weaver},
  {Zabalza}, {Edwards}, {Azalee Bostroem}, {Burke}, {Casey}, {Crawford},
  {Dencheva}, {Ely}, {Jenness}, {Labrie}, {Lim}, {Pierfederici}, {Pontzen},
  {Ptak}, {Refsdal}, {Servillat}, \& {Streicher}}]{2013A&A...558A..33A}
{Astropy Collaboration}, {Robitaille}, T.~P., {Tollerud}, E.~J., {et~al.} 2013,
  \href{http://dx.doi.org/10.1051/0004-6361/201322068}{\JournalTitle{\aap},
  558, A33}

\bibitem[{{Baraffe} {et~al.}(2003){Baraffe}, {Chabrier}, {Barman}, {Allard}, \&
  {Hauschildt}}]{2003A&A...402..701B}
{Baraffe}, I., {Chabrier}, G., {Barman}, T.~S., {Allard}, F., \& {Hauschildt},
  P.~H. 2003,
  \href{http://dx.doi.org/10.1051/0004-6361:20030252}{\JournalTitle{\aap}, 402,
  701}

\bibitem[{{Bihain} {et~al.}(2013){Bihain}, {Scholz}, {Storm}, \&
  {Schnurr}}]{2013A&A...557A..43B}
{Bihain}, G., {Scholz}, R.-D., {Storm}, J., \& {Schnurr}, O. 2013,
  \href{http://dx.doi.org/10.1051/0004-6361/201322141}{\JournalTitle{\aap},
  557, A43}

\bibitem[{{Blake} {et~al.}(2010){Blake}, {Charbonneau}, \&
  {White}}]{Blake:2010gf}
{Blake}, C.~H., {Charbonneau}, D., \& {White}, R.~J. 2010,
  \href{http://dx.doi.org/10.1088/0004-637X/723/1/684}{\JournalTitle{\apj},
  723, 684}

\bibitem[{{Burgasser} {et~al.}(2012){Burgasser}, {Luk}, {Dhital}, {Bardalez
  Gagliuffi}, {Nicholls}, {Prato}, {West}, \& {Lepine}}]{Burgasser:2012qy}
{Burgasser}, A.~J., {Luk}, C., {Dhital}, S., {et~al.} 2012, \JournalTitle{ArXiv
  e-prints}, \href{http://arxiv.org/abs/1208.0352}{{\sffamily arXiv:1208.0352
  [astro-ph.SR]}}

\bibitem[{{Burgasser} {et~al.}(2013){Burgasser}, {Sheppard}, \&
  {Luhman}}]{2013ApJ...772..129B}
{Burgasser}, A.~J., {Sheppard}, S.~S., \& {Luhman}, K.~L. 2013,
  \href{http://dx.doi.org/10.1088/0004-637X/772/2/129}{\JournalTitle{\apj},
  772, 129}

\bibitem[{{Burgasser} {et~al.}(2014){Burgasser}, {Gillon}, {Faherty},
  {Radigan}, {Triaud}, {Plavchan}, {Street}, {Jehin}, {Delrez}, \&
  {Opitom}}]{2014ApJ...785...48B}
{Burgasser}, A.~J., {Gillon}, M., {Faherty}, J.~K., {et~al.} 2014,
  \href{http://dx.doi.org/10.1088/0004-637X/785/1/48}{\JournalTitle{\apj}, 785,
  48}

\bibitem[{{Burgasser} {et~al.}(2015){Burgasser}, {Gillon}, {Melis}, {Bowler},
  {Michelsen}, {Bardalez Gagliuffi}, {Gelino}, {Jehin}, {Delrez}, {Manfroid},
  \& {Blake}}]{2015AJ....149..104B}
{Burgasser}, A.~J., {Gillon}, M., {Melis}, C., {et~al.} 2015,
  \href{http://dx.doi.org/10.1088/0004-6256/149/3/104}{\JournalTitle{\aj}, 149,
  104}

\bibitem[{{Burrows} {et~al.}(1997){Burrows}, {Marley}, {Hubbard}, {Lunine},
  {Guillot}, {Saumon}, {Freedman}, {Sudarsky}, \&
  {Sharp}}]{1997ApJ...491..856B}
{Burrows}, A., {Marley}, M., {Hubbard}, W.~B., {et~al.} 1997,
  \href{http://dx.doi.org/10.1086/305002}{\JournalTitle{\apj}, 491, 856}

\bibitem[{{Cushing} {et~al.}(2014){Cushing}, {Kirkpatrick}, {Gelino}, {Mace},
  {Skrutskie}, \& {Gould}}]{2014AJ....147..113C}
{Cushing}, M.~C., {Kirkpatrick}, J.~D., {Gelino}, C.~R., {et~al.} 2014,
  \href{http://dx.doi.org/10.1088/0004-6256/147/5/113}{\JournalTitle{\aj}, 147,
  113}

\bibitem[{{Cushing} {et~al.}(2006){Cushing}, {Roellig}, {Marley}, {Saumon},
  {Leggett}, {Kirkpatrick}, {Wilson}, {Sloan}, {Mainzer}, {Van Cleve}, \&
  {Houck}}]{Cushing:2006fk}
{Cushing}, M.~C., {Roellig}, T.~L., {Marley}, M.~S., {et~al.} 2006,
  \href{http://dx.doi.org/10.1086/505637}{\JournalTitle{\apj}, 648, 614}

\bibitem[{{Dupuy} \& {Liu}(2012)}]{Dupuy:2012fk}
{Dupuy}, T.~J., \& {Liu}, M.~C. 2012,
  \href{http://dx.doi.org/10.1088/0067-0049/201/2/19}{\JournalTitle{\apjs},
  201, 19}

\bibitem[{{Fischer} {et~al.}(2003){Fischer}, {Vrba}, {Toomey}, {Lucke}, {Wang},
  {Henden}, {Robichaud}, {Onaka}, {Hicks}, {Harris}, {Stahlberger},
  {Kosakowski}, {Dudley}, \& {Johnston}}]{Fischer:2003fj}
{Fischer}, J., {Vrba}, F.~J., {Toomey}, D.~W., {et~al.} 2003,
  \href{http://dx.doi.org/10.1117/12.461033}{in Society of Photo-Optical
  Instrumentation Engineers (SPIE) Conference Series, Vol. 4841, Society of
  Photo-Optical Instrumentation Engineers (SPIE) Conference Series, ed. {M.~Iye
  \& A.~F.~M.~Moorwood}}, 564

\bibitem[{{Gagn{\'e}} {et~al.}(2014){Gagn{\'e}}, {Lafreni{\`e}re}, {Doyon},
  {Malo}, \& {Artigau}}]{2014ApJ...783..121G}
{Gagn{\'e}}, J., {Lafreni{\`e}re}, D., {Doyon}, R., {Malo}, L., \& {Artigau},
  {\'E}. 2014,
  \href{http://dx.doi.org/10.1088/0004-637X/783/2/121}{\JournalTitle{\apj},
  783, 121}

\bibitem[{{Gizis} {et~al.}(2011){Gizis}, {Burgasser}, {Faherty}, {Castro}, \&
  {Shara}}]{2011AJ....142..171G}
{Gizis}, J.~E., {Burgasser}, A.~J., {Faherty}, J.~K., {Castro}, P.~J., \&
  {Shara}, M.~M. 2011,
  \href{http://dx.doi.org/10.1088/0004-6256/142/5/171}{\JournalTitle{\aj}, 142,
  171}

\bibitem[{{Golimowski} {et~al.}(2004){Golimowski}, {Leggett}, {Marley}, {Fan},
  {Geballe}, {Knapp}, {Vrba}, {Henden}, {Luginbuhl}, {Guetter}, {Munn},
  {Canzian}, {Zheng}, {Tsvetanov}, {Chiu}, {Glazebrook}, {Hoversten},
  {Schneider}, \& {Brinkmann}}]{2004AJ....127.3516G}
{Golimowski}, D.~A., {Leggett}, S.~K., {Marley}, M.~S., {et~al.} 2004,
  \JournalTitle{\aj}, 127, 3516

\bibitem[{{Hook} {et~al.}(2004){Hook}, {J{\o}rgensen}, {Allington-Smith},
  {Davies}, {Metcalfe}, {Murowinski}, \& {Crampton}}]{Hook:2004lr}
{Hook}, I.~M., {J{\o}rgensen}, I., {Allington-Smith}, J.~R., {et~al.} 2004,
  \href{http://dx.doi.org/10.1086/383624}{\JournalTitle{\pasp}, 116, 425}

\bibitem[{{Hu{\'e}lamo} {et~al.}(2015){Hu{\'e}lamo}, {Ivanov}, {Kurtev},
  {Girard}, {Borissova}, {Mawet}, {Mu{\v z}i{\'c}}, {C{\'a}ceres}, {Melo},
  {Sterzik}, \& {Minniti}}]{2015A&A...578A...1H}
{Hu{\'e}lamo}, N., {Ivanov}, V.~D., {Kurtev}, R., {et~al.} 2015,
  \href{http://dx.doi.org/10.1051/0004-6361/201525634}{\JournalTitle{\aap},
  578, A1}

\bibitem[{{Ivanov} {et~al.}(2015){Ivanov}, {Vaisanen}, {Kniazev}, {Beletsky},
  {Mamajek}, {Mu{\v z}i{\'c}}, {Beam{\'{\i}}n}, {Boffin}, {Pourbaix}, {Gandhi},
  {Gulbis}, {Monaco}, {Saviane}, {Kurtev}, {Mawet}, {Borissova}, \&
  {Minniti}}]{2015A&A...574A..64I}
{Ivanov}, V.~D., {Vaisanen}, P., {Kniazev}, A.~Y., {et~al.} 2015,
  \href{http://dx.doi.org/10.1051/0004-6361/201424883}{\JournalTitle{\aap},
  574, A64}

\bibitem[{{Kenyon} \& {Hartmann}(1995)}]{1995ApJS..101..117K}
{Kenyon}, S.~J., \& {Hartmann}, L. 1995,
  \href{http://dx.doi.org/10.1086/192235}{\JournalTitle{\apjs}, 101, 117}

\bibitem[{{Khandrika} {et~al.}(2013){Khandrika}, {Burgasser}, {Melis}, {Luk},
  {Bowsher}, \& {Swift}}]{Khandrika:2013fk}
{Khandrika}, H., {Burgasser}, A.~J., {Melis}, C., {et~al.} 2013,
  \href{http://dx.doi.org/10.1088/0004-6256/145/3/71}{\JournalTitle{\aj}, 145,
  71}

\bibitem[{{Kirkpatrick} {et~al.}(1999){Kirkpatrick}, {Reid}, {Liebert},
  {Cutri}, {Nelson}, {Beichman}, {Dahn}, {Monet}, {Gizis}, \&
  {Skrutskie}}]{1999ApJ...519..802K}
{Kirkpatrick}, J.~D., {Reid}, I.~N., {Liebert}, J., {et~al.} 1999,
  \href{http://dx.doi.org/10.1086/307414}{\JournalTitle{\apj}, 519, 802}

\bibitem[{{Kirkpatrick} {et~al.}(2012){Kirkpatrick}, {Gelino}, {Cushing},
  {Mace}, {Griffith}, {Skrutskie}, {Marsh}, {Wright}, {Eisenhardt}, {McLean},
  {Mainzer}, {Burgasser}, {Tinney}, {Parker}, \&
  {Salter}}]{2012ApJ...753..156K}
{Kirkpatrick}, J.~D., {Gelino}, C.~R., {Cushing}, M.~C., {et~al.} 2012,
  \href{http://dx.doi.org/10.1088/0004-637X/753/2/156}{\JournalTitle{\apj},
  753, 156}

\bibitem[{{Kniazev} {et~al.}(2013){Kniazev}, {Vaisanen}, {Mu{\v z}i{\'c}},
  {Mehner}, {Boffin}, {Kurtev}, {Melo}, {Ivanov}, {Girard}, {Mawet},
  {Schmidtobreick}, {Huelamo}, {Borissova}, {Minniti}, {Ishibashi}, {Potter},
  {Beletsky}, {Buckley}, {Crawford}, {Gulbis}, {Kotze}, {Miszalski},
  {Pickering}, {Romero Colmenero}, \& {Williams}}]{2013ApJ...770..124K}
{Kniazev}, A.~Y., {Vaisanen}, P., {Mu{\v z}i{\'c}}, K., {et~al.} 2013,
  \href{http://dx.doi.org/10.1088/0004-637X/770/2/124}{\JournalTitle{\apj},
  770, 124}

\bibitem[{{Livingston} \& {Wallace}(1991)}]{Livingston:1991fj}
{Livingston}, W., \& {Wallace}, L. 1991, {An atlas of the solar spectrum in the
  infrared from 1850 to 9000 cm-1 (1.1 to 5.4 micrometer)} (National Solar
  Observatory)

\bibitem[{{Luhman}(2013)}]{2013ApJ...767L...1L}
{Luhman}, K.~L. 2013,
  \href{http://dx.doi.org/10.1088/2041-8205/767/1/L1}{\JournalTitle{\apjl},
  767, L1}

\bibitem[{{Luhman}(2014)}]{2014ApJ...786L..18L}
---. 2014,
  \href{http://dx.doi.org/10.1088/2041-8205/786/2/L18}{\JournalTitle{\apjl},
  786, L18}

\bibitem[{{Magazzu} {et~al.}(1993){Magazzu}, {Martin}, \&
  {Rebolo}}]{1993ApJ...404L..17M}
{Magazzu}, A., {Martin}, E.~L., \& {Rebolo}, R. 1993,
  \href{http://dx.doi.org/10.1086/186733}{\JournalTitle{\apjl}, 404, L17}

\bibitem[{{Mamajek} {et~al.}(2015){Mamajek}, {Barenfeld}, {Ivanov}, {Kniazev},
  {V{\"a}is{\"a}nen}, {Beletsky}, \& {Boffin}}]{2015ApJ...800L..17M}
{Mamajek}, E.~E., {Barenfeld}, S.~A., {Ivanov}, V.~D., {et~al.} 2015,
  \href{http://dx.doi.org/10.1088/2041-8205/800/1/L17}{\JournalTitle{\apjl},
  800, L17}

\bibitem[{{McLean} {et~al.}(2000){McLean}, {Graham}, {Becklin}, {Figer},
  {Larkin}, {Levenson}, \& {Teplitz}}]{McLean:2000lr}
{McLean}, I.~S., {Graham}, J.~R., {Becklin}, E.~E., {et~al.} 2000, in Society
  of Photo-Optical Instrumentation Engineers (SPIE) Conference Series, Vol.
  4008, Society of Photo-Optical Instrumentation Engineers (SPIE) Conference
  Series, ed. M.~{Iye} \& A.~F. {Moorwood}, 1048

\bibitem[{{McLean} {et~al.}(2007){McLean}, {Prato}, {McGovern}, {Burgasser},
  {Kirkpatrick}, {Rice}, \& {Kim}}]{2007ApJ...658.1217M}
{McLean}, I.~S., {Prato}, L., {McGovern}, M.~R., {et~al.} 2007,
  \href{http://dx.doi.org/10.1086/511740}{\JournalTitle{\apj}, 658, 1217}

\bibitem[{{P{\'e}rez Garrido} {et~al.}(2014){P{\'e}rez Garrido}, {Lodieu},
  {B{\'e}jar}, {Ruiz}, {Gauza}, {Rebolo}, \& {Zapatero
  Osorio}}]{2014A&A...567A...6P}
{P{\'e}rez Garrido}, A., {Lodieu}, N., {B{\'e}jar}, V.~J.~S., {et~al.} 2014,
  \href{http://dx.doi.org/10.1051/0004-6361/201423615}{\JournalTitle{\aap},
  567, A6}

\bibitem[{{Radigan} {et~al.}(2012){Radigan}, {Jayawardhana}, {Lafreni{\`e}re},
  {Artigau}, {Marley}, \& {Saumon}}]{Radigan:2012vn}
{Radigan}, J., {Jayawardhana}, R., {Lafreni{\`e}re}, D., {et~al.} 2012,
  \href{http://dx.doi.org/10.1088/0004-637X/750/2/105}{\JournalTitle{\apj},
  750, 105}

\bibitem[{{Saumon} \& {Marley}(2008)}]{2008ApJ...689.1327S}
{Saumon}, D., \& {Marley}, M.~S. 2008,
  \href{http://dx.doi.org/10.1086/592734}{\JournalTitle{\apj}, 689, 1327}

\bibitem[{{Scholz}(2014)}]{2014A&A...561A.113S}
{Scholz}, R.-D. 2014,
  \href{http://dx.doi.org/10.1051/0004-6361/201323015}{\JournalTitle{\aap},
  561, A113}

\bibitem[{{Skrutskie} {et~al.}(2006){Skrutskie}, {Cutri}, {Stiening},
  {Weinberg}, {Schneider}, {Carpenter}, {Beichman}, {Capps}, {Chester},
  {Elias}, {Huchra}, {Liebert}, {Lonsdale}, {Monet}, {Price}, {Seitzer},
  {Jarrett}, {Kirkpatrick}, {Gizis}, {Howard}, {Evans}, {Fowler}, {Fullmer},
  {Hurt}, {Light}, {Kopan}, {Marsh}, {McCallon}, {Tam}, {Van Dyk}, \&
  {Wheelock}}]{2mass}
{Skrutskie}, M.~F., {Cutri}, R.~M., {Stiening}, R., {et~al.} 2006,
  \href{http://dx.doi.org/10.1086/498708}{\JournalTitle{\aj}, 131, 1163}

\bibitem[{{Vrba} {et~al.}(2004){Vrba}, {Henden}, {Luginbuhl}, {Guetter},
  {Munn}, {Canzian}, {Burgasser}, {Kirkpatrick}, {Fan}, {Geballe},
  {Golimowski}, {Knapp}, {Leggett}, {Schneider}, \&
  {Brinkmann}}]{2004AJ....127.2948V}
{Vrba}, F.~J., {Henden}, A.~A., {Luginbuhl}, C.~B., {et~al.} 2004,
  \JournalTitle{\aj}, 127, 2948

\bibitem[{{Wright} {et~al.}(2010){Wright}, {Eisenhardt}, {Mainzer}, {Ressler},
  {Cutri}, {Jarrett}, {Kirkpatrick}, {Padgett}, {McMillan}, {Skrutskie},
  {Stanford}, {Cohen}, {Walker}, {Mather}, {Leisawitz}, {Gautier}, {McLean},
  {Benford}, {Lonsdale}, {Blain}, {Mendez}, {Irace}, {Duval}, {Liu}, {Royer},
  {Heinrichsen}, {Howard}, {Shannon}, {Kendall}, {Walsh}, {Larsen}, {Cardon},
  {Schick}, {Schwalm}, {Abid}, {Fabinsky}, {Naes}, \&
  {Tsai}}]{2010AJ....140.1868W}
{Wright}, E.~L., {Eisenhardt}, P.~R.~M., {Mainzer}, A.~K., {et~al.} 2010,
  \href{http://dx.doi.org/10.1088/0004-6256/140/6/1868}{\JournalTitle{\aj},
  140, 1868}

\end{thebibliography}


\end{document}